\documentclass[12pt]{article}
\pdfoutput=1
\usepackage[utf8]{inputenc}
\usepackage[T1]{fontenc}
\usepackage{float}
\usepackage{authblk}
\usepackage[normalem]{ulem}
\usepackage{xcolor}
\usepackage{amsmath}
\usepackage{amsthm}
\usepackage{amssymb}
\usepackage{amsfonts}
\usepackage{adjustbox}
\usepackage{bbm}
\usepackage{graphicx}
\usepackage{booktabs}
\usepackage{mathtools}
\usepackage{multirow,array}
\usepackage{fullpage}
\usepackage{comment}
\usepackage[mathlines]{lineno}
\usepackage{multirow}
\usepackage{subcaption}
\usepackage{bm}
\usepackage{caption}

\usepackage[colorlinks = true,
            linkcolor = teal,
            urlcolor  = teal,
            citecolor = teal]{hyperref}

\title{Temporal dynamics of goal scoring in soccer}

\author[1]{Guteraa~Ayana\thanks{Contributed equally as first authors.}}
\author[1]{Alexander~Ehlert$^*$}
\author[2]{Joseph~Ehlert}
\author[2]{Luca~Santagata}
\author[2]{Maddalena~Torricelli}
\author[2]{Brennan~Klein}

\affil[1]{Breck School, Golden Valley, MN}
\affil[2]{Network Science Institute, Northeastern University, Boston, MA}

\begin{document}
\maketitle
% \linenumbers
\pagenumbering{arabic}

\begin{abstract}
We investigated the temporal distribution of goals in soccer using event-level data from 3,433 matches across 21 leagues and competitions. Contrary to the prevailing notion of randomness, we found that the probability of a goal being scored is higher as matches progress, and we observed fewer-than-expected goals in the early minutes of each half. Further analysis of the time \textit{between} subsequent goals shows an exponential decay, indicating that most goals naturally cluster closer together in time. By splitting this distribution by the team that scores the next goal, we observe bursty goal-scoring dynamics, wherein the same team is more likely to score again shortly after its previous goal. These findings highlight the importance of match context---whether driven by fatigue, tactical adaptations, or psychological momentum---in shaping when teams are able to score. Moreover, the results open avenues for extending data-driven methods for identifying high-impact moments in a match and refining strategic decision-making in soccer's evolving analytical landscape.
\end{abstract}

\section{Introduction}\label{sec:intro}

Soccer, also known as football throughout most of the world, is a sport played between two teams of 11 players on a rectangular field with a goal at each end. Soccer is governed by the rules of the game known as the Laws of the Game, established by the International Football Association Board \cite{ifab2025}. The objective of the game is to score goals by maneuvering a ball into the opposing team's goal using any part of the body except the hands and arms, with the exception of goalkeepers who are allowed to use their hands within a designated box. Matches are divided into two halves of 45 minutes each, with a brief halftime interval. The half length can be extended to account for stoppage time coming from player injuries, substitutions, and more. With its simplicity and minimal equipment requirements, soccer has grown into the most widely played and watched sport in the world \cite{scottishsun2024}. 

Analytics plays an increasingly prominent role in contemporary professional soccer, influencing team tactics, player contracts, and coaching strategies. Before this popularization of data-driven methodologies, soccer tactics were predominantly shaped by intuition, experience, and observational analysis \cite{gonzalez2023evolution, rein2016big}. Statistical modeling and machine learning algorithms are now widely employed to analyze player performance, predict match outcomes, and optimize game plans \cite{mark2024liverpool}. Today, the average television viewer of professional soccer encounters substantially more statistical information, with metrics like the Expected Goals (xG), pass success rates, and heat maps now commonly used to summarize key patterns in a game.

This growing understanding of the role of data in professional soccer has coincided with a new range of tactical innovations and trends. Amid this heterogeneity, we also see a renewed interest in uncovering new ways of quantifying and characterizing \textit{commonalities} between teams. In this work, we document gameplay trends that persist across countries, leagues, and seasons. Specifically, we measure the time that goals are scored in games and---in games with two or more goals---the time difference between subsequent goals.

Drawing on a large dataset of professional matches across multiple leagues and competitions, our analyses focus on identifying distinct moments within games that may elevate or reduce the likelihood of scoring, relative to a null model of random goal timing. By contrasting observed events against this theoretical framework, we find when teams are most/least likely to score a goal. These observations motivate a deeper look into whether late-match attacking pushes, early-half lulls, or other situational shifts can decisively influence goal patterns. Ultimately, we aim to resolve the question: \emph{Is there a random distribution of when goals are scored?}

\section{Data \& Methods}\label{sec:data_methods}

\subsection{Data Source: Hudl StatsBomb}\label{sec:data_statsbomb}
The dataset used in this study consists of event information collected from 3,433 games across 21 different leagues and competitions, including: FIFA World Cup, African Cup of Nations, Premier League, 1. Bundesliga, La Liga, Serie A, Champions League, Major League Soccer (MLS), and National Women's Soccer League (NWSL). These leagues and competitions span various seasons, and are aggregated and stored by StatsBomb. The basic unit of this data is an \textit{event}, or actions that happen over the course of a game, including elements such as passes, carries, shots, substitutions, and fouls. This dataset has three partitions: events, lineups, and matches. Detailed information and schema for the StatsBomb data as well as available leagues and competitions can be found at \url{https://github.com/statsbomb/open-data}.

\paragraph{Events:}
Each game has a JSON file out of the 3,433 files holding the explicit game data, identified by the match ID. Each game file consists of an array with every event that was recorded in the game as a series of discreet, timestamped events. Each event contains 20 descriptive attributes, including the event timestamp, the type of event (i.e., goal, pass, dribble, etc.), and the players involved. In sum, there are 34 different event types, with ``Ball Receipt'', ``Shot'', and ``Pass'' as the most common. The ``Shot'' event contains additional information on the shot event, such as the start/end location of the shot, the xG (a measure of expectation to score), and the shot's outcome (e.g.~``Goal'', ``Blocked'', ``Saved'', etc. See Fig.~\ref{fig:Three_Schematics}a).

\paragraph{Lineups:}
\emph{Lineups} contains information on the lineups and lineup changes for the same 3,433 games. Each file consists of five elements for each player: ``ID'' (generated by StatsBomb specific to the player), ``name'', ``nickname'', ``jersey number'', and ``country/nationality'' (See Fig.~\ref{fig:Three_Schematics}b).

\paragraph{Matches:}
\emph{Matches} contains the specific league or competition (e.g., Premier League) and season (e.g., 2015/2016) for each of the 3,433 matches. It is split into leagues, which then have season inside (See Fig.~\ref{fig:Three_Schematics}c). Each section contains various amounts of JSON files, with each file holding information on matches played during a specific season. The matches folder was used to collect all of the 3,433 match ID's, and to collect specific match IDs from a league and/or season.

\subsection{A null model for goal scoring} \label{sec:methods_time_between}
In order to compare the observed distribution of the timing of (and between) goals, we introduce a null model for estimating the expected number of goals within any given time interval. To construct this goal-scoring null model, we assume that the probability of scoring a goal is consistent across all minutes and matches, defined by the average number of goals per minute, $\bar{G}$:
\begin{equation}\label{eq:goals-per-min} 
    \bar{G} = \frac{1}{N} \sum_{i=1}^N \frac{G_i}{T_i}, 
\end{equation}
where $G_i$, $T_i$, and $N$ represent the number of goals, the duration in minutes of match $i$, and the number of matches, respectively. We simulated 6,000,000 games with goals scored at a rate of $\bar{G}$ and game lengths set to the average game length, $\bar{T}$. In each simulated game, goals were randomly assigned to one of the two teams.

\section{Results}\label{sec:results}

\subsection{Goal timing} \label{sec:results_goal_timing}

On first glance, there does not seem to be an intuitive rationale for why we would expect a goal to be scored at any specific time during a game. This suggests that, by measuring the time that goals are scored across many games, we would expect to see a maximum entropy, uniform distribution across the duration of a game. In Fig.~\ref{fig:results-time-between}a, we show that this is not the case. Instead, we see that the probability of a goal being scored \textit{increases} as the game progresses.

We propose that goal scoring exhibits non-random temporal patterns influenced by specific game dynamics. Specifically, we hypothesize that there are particular intervals---such as the final 10 minutes of each half or immediately following a substitution---where the frequency of goals significantly differs from the uniform expectation. By identifying these periods, we aim to uncover strategic or psychological factors that contribute to heightened or depressed goal-scoring activity.

To test this hypothesis, we start with the assumption that the aggregation of goals from many matches should be randomly and uniformly distributed across the minutes in a game, looking for deviations from expectation. Early in the game, we find that goals occur less frequently compared to the expected frequency, and as the game progresses, we find that goals are more likely to be scored (See Fig.~\ref{fig:results-time-between}a, $r\approx0.438$, $p\approx7.13\times10^{-6}$). This upward trend could be attributed to various factors, such as substitutions, defensive player fatigue, or tactical shifts. Additionally, the actual frequency also appears to drop between 45 and 50 minutes, likely because of the halftime restart. We further quantify these deviations from a uniform distribution by performing a $\chi^2$-test, finding a significant difference ($\chi^2 \approx 288.62, p=3.72\times10^{-21}$). We therefore conclude that the observed distribution deviates significantly from what would be expected under a uniform distribution, demonstrating the existence of time intervals when more goals are scored than expected.

\subsubsection{Time between goals shows exponential decay}
While the observed timing of goals throughout a game is not uniformly random, this does not immediately reveal the strategic or performance factors that affect when a goal is scored. As such, we next investigate whether the timing of a goal is dependent on other goals scored. Pundits frequently comment that the five minutes after a goal are the most dangerous for additional goals. We therefore hypothesize that teams are most likely to score a goal shortly after another goal was scored, leading to an average short time between goals.

\begin{figure}
    \centering
    \includegraphics[width=\textwidth]{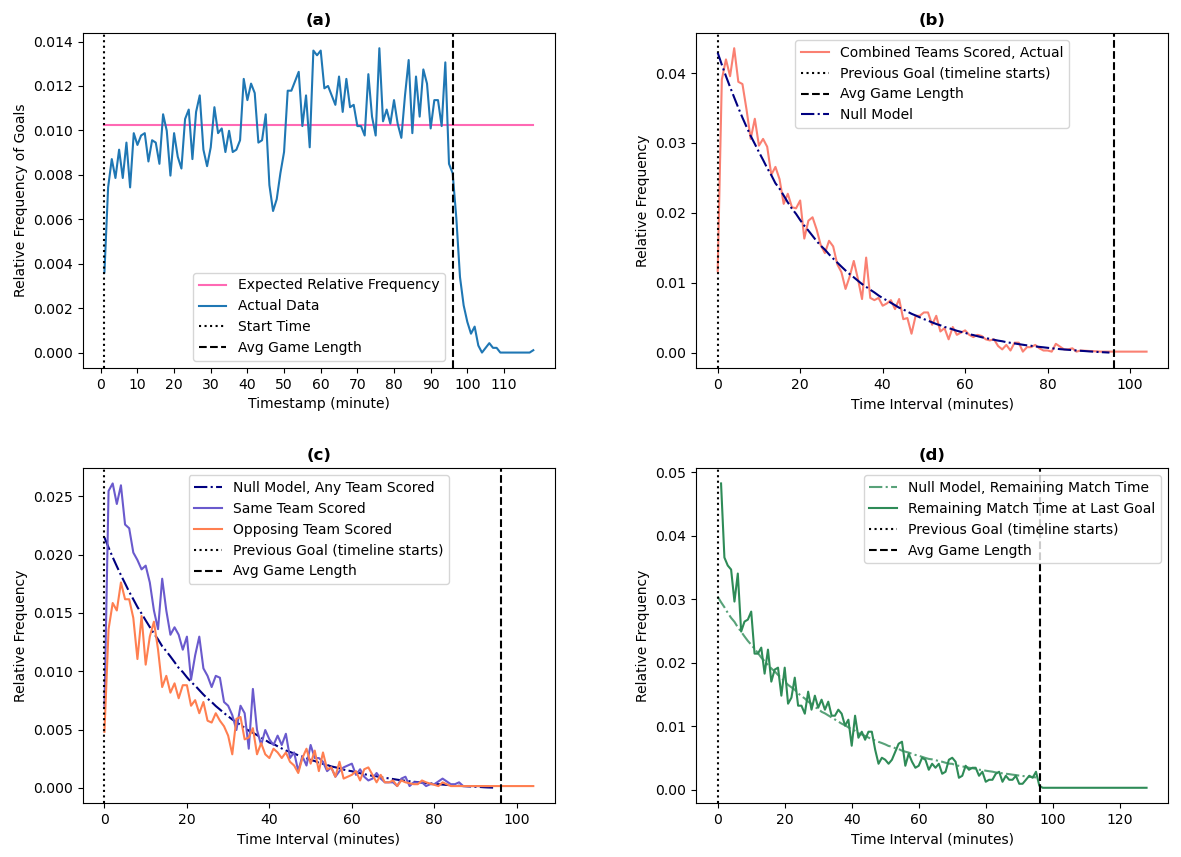}
    \captionsetup{font=footnotesize}
    \caption{\textbf{Temporal analysis of goal-scoring.} \textbf{(a)} Comparison of observed relative frequency (blue) and expected goal distribution (pink) across match duration. We see that the actual frequency of goals varies throughout the match, oscillating around expected frequency. The frequency trends upwards as the match progresses, until the average match length, where the frequency of goals drops sharply.
    \textbf{(b)} Comparison of actual data (salmon) and null model (blue) for combined time intervals between goals. The vertical line marks the moment the previous goal was scored and serves as the starting reference (t=0) for measuring the intervals to subsequent goals. The actual data appears to very closely follow the null model. The shorter time intervals have a higher relative frequency of goals scored, therefore, there is a time frame---right after a goal is scored---where more goals are scored compared to other time frames.
    \textbf{(c)} Comparison of actual data (solid) and null model (dash-dot) for same team (blue) and opposing team (orange). The frequency is normalized across both event types, i.e. same team scoring and opposing team scoring. For the null model (See Section~\ref{sec:methods_time_between}), the same team scored and opposing team scored lines are identical, so only one is shown. The null model, similar to the actual data,  exponentially decays. The actual data appears to closely follow the null model at longer time intervals, but at shorter time intervals, there is a large difference between the actual data and the null model for time intervals between 0 and roughly 30 minutes. Therefore, for the same team, immediately after the goal, they are more likely to score than expected, and for the opposing team, immediately after the goal, they are less likely to score than expected.
    \textbf{(d)} Comparison of actual data (solid) and null model (dash-dot) for the remaining match time at the last goal. The observed frequencies closely follow the null model for most time intervals, with the exception of the shortest time intervals, which suggests that teams are more likely to score the last goal of a match closer to the end of a match than would be expected.}
    \label{fig:results-time-between}
\end{figure}

To test this hypothesis, we subset our dataset into games that had at least two goals. Among these games, we measure the time difference between goals scored, and plot the distribution of these time differences (Fig.~\ref{fig:results-time-between}b). We also calculated the time interval between goals scored in the games simulated in the previous section to obtain an appropriate null model. Strikingly, many goals are indeed scored within several minutes of the previous goal scored, which can be fit with an exponential decay (Fig.~\ref{fig:results-time-between}b). The null model shows a nearly identical pattern, suggesting that even though most goals are scored shortly after another goal, this is likely due to time-dependent constraints rather than the mentality of the players or the cadence of the match changing following a goal. To test if the difference between the observed time intervals and null model was significant, we conducted a $\chi^2$-test ($\chi^2 \approx 0.044$, $p=1$), suggesting that goals are well explained by the null model. This result implies that there are many more goals scored in the five minutes after a goal, but this is, in fact, explained by the average rate of goal scoring.

\subsubsection{Teams exhibit bursty goal-scoring behavior after scoring}
To further explore differences between our null model for goal-scoring and the observed data, we separate the data by the team that scores. This lets us highlight any scoring dynamics that might remain hidden when both teams are aggregated. This hypothesis is drawn from the idea that although the overall goal scoring rate will not change, the rate for each team may change upon scoring a goal.

To determine whether previous goals affect the timing of the subsequent goal, we explored the time interval between goals scored by the team that previously scored and the team that previously conceded. If the same team scores twice in a row, this time interval is classified as same team and vice versa for the opposing team scoring next. Predictably, same team and opposing team time intervals both exhibit an exponential decay similar to what was observed in the aggregated distribution (Fig.~\ref{fig:results-time-between}c). The same team scores the next goal more often than the opposing team across all time intervals, suggesting \textit{bursty} goal-scoring behavior, a common phenomena in humans \cite{Barabasi2005bursts, karsai2024measuringmodelingburstyhuman}. This marked difference could be attributed to the fact that better teams are more likely to score \textit{in general}, and as such, they are more likely to score consecutive goals. It could also be a result of hot streaks on the team, where team confidence increases with a goal, or the conceding team loses confidence.

Finally, we compared the observed time remaining in the match when the last goal is scored to our null model, finding that teams are more likely to score the last goal of a match closer to the end of a match than expected, evidenced by the observed time remaining far exceeding the expected amount of time left in the game (Fig.~\ref{fig:results-time-between}d). Furthermore, the same team scores the next goal more frequently than expected, and the opposing team scores the next goal less than expected, indicating that goal scoring for the same team is well described as bursty behavior.

\section{Discussion \& Conclusion}
The results presented here suggest that goals in soccer do not follow a random, uniform temporal distribution. Instead, we observe that specific intervals in a match significantly deviate from the expected goal-scoring rate under a uniform time distribution. One particularly salient trend is the tendency for goal frequency to be lower at the beginning of each half, with a marked increase in the later stages. Potential factors driving this pattern include tactical adjustments once teams realize they need to push for a goal, fatigue or inattention in defensive shape, and strategic substitutions that can provide fresh legs or shift momentum.

Closely tied to this finding is the phenomenon of time between goals, in which teams that previously scored are more likely to score a consecutive goal sooner than expected relative to an entirely random process. Though the time between pairs of goals can be modeled well using a simple exponential-decay null model, splitting goals into same team and opposing team categories reveals temporally concentrated goal scoring, favoring the team that just scored. Such burstiness is well-documented in various human activities, where a success or ``hot streak'' appears to positively reinforce subsequent success \cite{Barabasi2005bursts, karsai2024measuringmodelingburstyhuman}. In soccer, this could reflect a variety of psychological or tactical mechanisms---for example, increased confidence and aggression after scoring, or conversely, disarray and demoralization in the conceding team. At the same time, better teams are more likely to score in general, and thus repeated scoring by the same side can be partly attributed to inherent skill differentials. This idea could be specifically incorporated into the null model by biasing the probability that a given team will score, on average. Regardless of the underlying cause, these results are a clear example of time-dependent effects on goal scoring.

Moreover, the findings reinforce our overarching hypothesis: there are distinct time frames during a soccer match when goals are more likely to occur. By comparing observed frequencies of goals scored to those expected under a uniform distribution, we show that not only do these heightened windows of opportunity exist---particularly shortly after a goal---but are significantly different than what a null model would suggest. This discovery lends strong support to the idea that scoring is context-dependent, reflecting changes in momentum, player fatigue, and tactical shifts as the clock winds down, implying that there are specific tactical or player adjustments that could be made to improve chances of scoring.

Our analysis establishes that soccer goals are not uniformly distributed over match time; rather, certain time frames consistently emerge as being particularly conducive to scoring. The early phases of each half show lower-than-expected goal frequency, whereas later phases and moments immediately following a goal scored exhibit elevated scoring likelihood. 

While this study used a large dataset of soccer matches, the specific sample we have access to is likely biased, as it is based on the data that is made publicly available from StatsBomb. Because of this, the measures observed might not be representative of professional soccer games in all leagues. In the future, these analyses should be redone with a larger, more systematic sample of games to better generalize the results found.

While our analysis primarily focused on uncovering these goal-timing nuances, our findings resonate with the broader literature exploring how being slightly behind can sometimes serve as a catalyst for improved performance. In the study titled ``Can Losing Lead to Winning?''---which examined collegiate and professional basketball games---researchers reported that teams trailing by a small margin at halftime had a higher-than-expected chance of eventually winning the game \cite{Berger2011losing}. This outcome was attributed to heightened motivation or effort induced by the psychological aversion to ``losing'' at the midpoint.

An intriguing avenue for future work is to test whether a similar pattern is observed in soccer. Given soccer's inherently low-scoring nature, this effect is more challenging to detect than in basketball, which may require new methodologies to identify a similar pattern. Does a team knowing that they are behind in a game (e.g., losing 0–1) spark an uptick in effort, focus, or attacking intensity that improves the trailing team's chances of leveling or taking the lead? The lower scoring rate in soccer complicates direct comparisons because fewer goals are scored overall, and a single goal can drastically alter match dynamics, mindsets, and the final result. Future research could explore how teams' in-game decision-making---such as defensive tactics or personnel substitutions---shifts when they are marginally trailing. Moreover, analyzing whether the observed bursty scoring patterns might be related to motivational surges akin to those described in the basketball literature would help determine if losing can lead to winning in low-scoring games like soccer.

Another avenue for future work is a deeper investigation of the specific events that occur in the build-up to goals. Instead of simply looking at when goals happen, further analyses can compare the possession sequences preceding a goal to identify the actions---such as incisive passes, successful dribbles, or off-the-ball movements---that tend to precede goals more frequently. Such an analysis could yield valuable tactical insights. For instance, if certain play patterns repeatedly surface before a goal is scored, coaches and analysts could focus on recreating or preempting these patterns during training. Uncovering these critical events might also shed light on how mental and physical conditions (e.g., player fatigue, the psychological boost of being slightly behind) interact with team strategies to catalyze goal-scoring opportunities. By systematically mapping out how and why these events cluster together prior to a goal, we can deepen our understanding of the game's dynamic flow and potentially enhance predictive models that forecast the likelihood of a goal under varying conditions. 

% Analysis of networks such as substitution networks, goal networks, and event networks offers another outlet to explore. Substitution networks have the potential to reveal that as new players enter the match, the event amounts and possession length could shift. Goal networks can help analyze pathways leading to goals, identifying key contributors and the sequence of events that maximize scoring potential. Event networks, which map relationships between passes, carries, pressures, and other in-game actions, can provide granular insights into how specific event chains influence game outcomes. These network-based analyses might provide insight on the roles of central players, distinguish critical transitions, and support coaching strategies by pinpointing optimal pathways and sequences for effective play. 

\subsection*{Additional information}
\paragraph{Data availability:} Data used in this work were downloaded from StatsBomb's free data portal at \url{https://github.com/statsbomb/open-data}.

\paragraph{Code availability:} Code to reproduce the analyses in this work will be available upon publication.

\paragraph{Acknowledgments:} The authors thank %Maddalenna Torricelli and 
Alessandro Vespignani for insightful discussions and constructive feedback.

\begin{sloppypar}
\bibliographystyle{unsrt}
\bibliography{main}
\end{sloppypar}

\clearpage

\appendix
\setcounter{figure}{0}
\setcounter{table}{0}
\setcounter{equation}{0}
\renewcommand\thefigure{\thesection.\arabic{figure}}
\renewcommand\thetable{\thesection.\arabic{table}}
\renewcommand\theequation{\thesection .\arabic{equation}}
% \begin{refsection}

\section{Supplementary Information}\label{sec:appendix_A}

\begin{figure}[h!]
    \centering
    \includegraphics[width=0.5\textwidth]{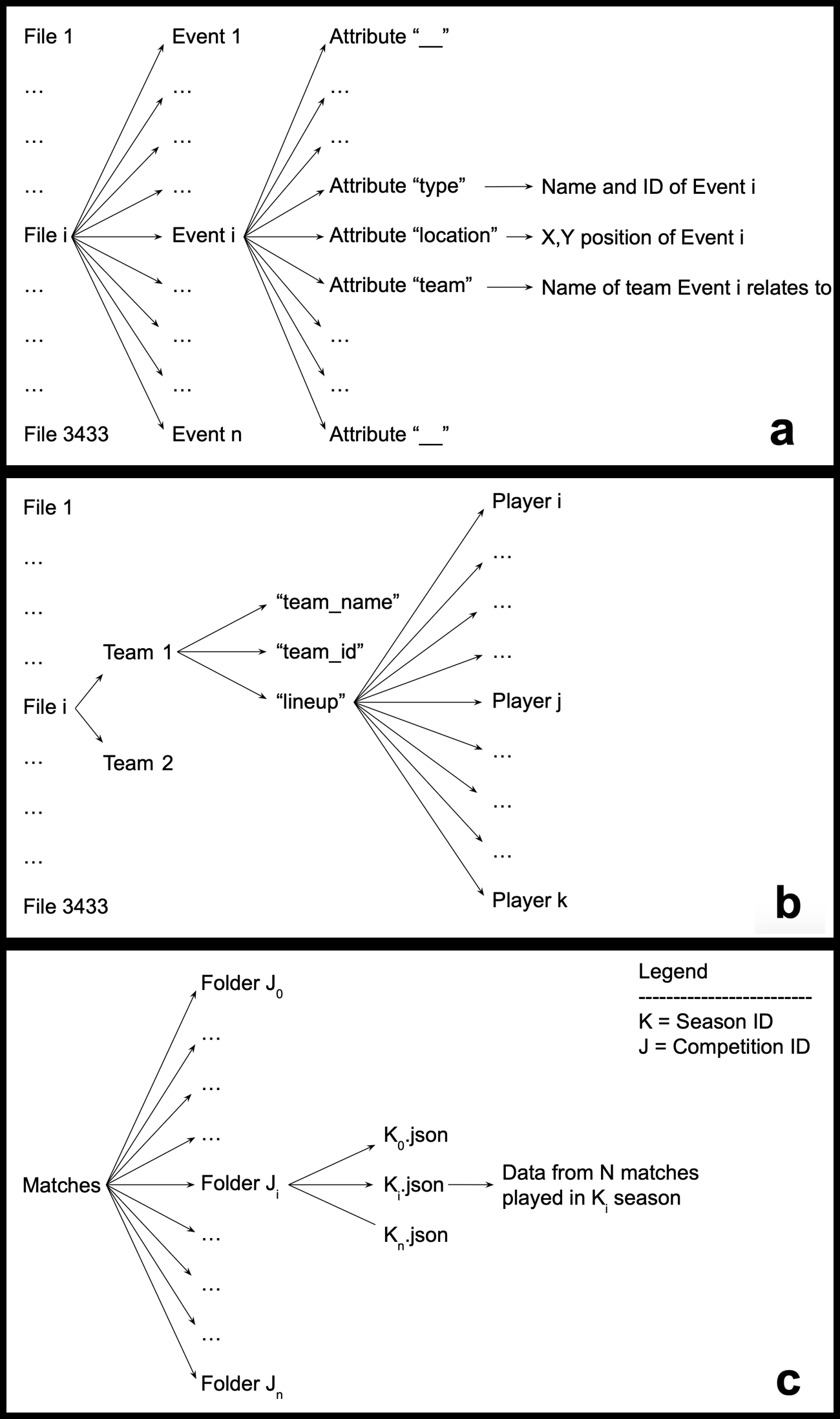}
    \caption{\textbf{(a)} Hierarchical representation of event data within each match file. Each file contains a series of events, labeled as Event 1 through Event n, each defined by 20 attributes such as \emph{type} (providing the name and ID of the event), \emph{location} (defining the X, Y position relative to the pitch), and \emph{team} (identifying the associated team). \textbf{(b)} Visualization of team data structure within match files. Each file includes data for two teams, identified as Team 1 and Team 2, with each team represented by attributes such as \emph{team\_name}, \emph{team\_id}, and \emph{lineup}. The \emph{lineup} connects each team to its players, represented individually. Not displayed, but within each Player object, there are five child attributes being: \emph{name}, \emph{nickname}, \emph{ID}, \emph{country}, and \emph{jersey number}. \textbf{(c)} Directory organization of match data across competitions and seasons. Matches are grouped into folders based on competition IDs (e.g., $J_0$, $J_1$,..., $J_n$), with each folder containing season-specific files labeled as $K_0$.json, $K_1$.json,..., $K_n$.json.}
    \label{fig:Three_Schematics}
\end{figure}

\end{document}